\documentclass[aps,prd,showpacs,twocolumn]{revtex4-1}

\usepackage{graphicx}
\usepackage{enumerate}
\usepackage{multirow}

\def\Journal#1#2#3#4{{#1} {{\bf #2},} {#4} {(#3)}}

\def\PLB{{Phys. Lett.}  B}
\def\PL{{Phys. Lett.}}

\def\PRL{ Phys. Rev. Lett.}

\def\PRD{{Phys. Rev.} D}
\def\PRC{{Phys. Rev.} C}

\def\ZPC{{Z. Phys. C}}

\def\EPJC{{Eur. Phys. J.} C}

\def\JHEP{{J. High Energy Phys.}}
\def\JPG{{J. Phys. G.}}

\def\IJMPA{{Int. J. Mod. Phys. A}}

\def\ra{\rightarrow}

\def\be{\begin{equation}}
\def\ee{\end{equation}}
\def\bea{\begin{eqnarray}}
\def\eea{\end{eqnarray}}

\def\etap{\eta^\prime}

\usepackage{relsize}
\def\babar{\mbox{\slshape B\kern-0.1em{\smaller A}\kern-0.1em  B\kern-0.1em{\smaller A\kern-0.2em R}}}

\begin{document}

%\preprint{}

\title{Determination of the $\eta$-$\eta^\prime$ mixing angle}

\author{Fu-Guang Cao}

\affiliation{Institute of Fundamental Sciences, Massey University, Private Bag 11 222, Palmerston North, New Zealand}

\date{\today}

\begin{abstract}
We extract $\eta$-$\eta^\prime$ mixing angle and the ratios of decay constants of light pseudoscalar mesons $\pi^0$, $\eta$ and $\etap$
using recently available \babar\ measurements on
$\eta$-photon and $\eta^\prime$-photon transition form factors and more accurate experimental data for the masses and two-photon decay widths
of the light pseduoscalar mesons. 
\end{abstract}

%\keywords{}

\maketitle

%\section{Introduction}
Determining the composition of $\eta$ and $\etap$ mesons attracted continuous interest in hadronic physics. 
The idea of $\eta$ and $\etap$ containing gluonic and intrinsic $|c\bar c \rangle$ components has long been employed in explaining many 
experimental results, including recent observations of large branching ratios for some decay processes of $J/\psi$ and $B$ mesons
into pseudoscalar mesons \cite{decays}.

There are three charge neutral states in the nonet of pseudoscalar mesons in SU(3)$_F$ quark model: 
$\pi^0$, $\eta_8$ and $\eta_1$.  The latter two mix to give the physical particles $\eta$ and $\eta^\prime$,
\bea
\left(
	\begin{array}{c}
	\eta \\
	\eta^\prime
	\end{array}
\right)
=\left(
	\begin{array}{cc}
	{\rm cos} \, \theta & -{\rm sin} \, \theta \\
	{\rm sin} \, \theta & {\rm cos} \, \theta
	\end{array}
\right)
\left(
	\begin{array}{c}
	\eta_8 \\
	\eta_1
	\end{array}
\right).
\label{eq:eta81basis}
\eea
Alternatively, one could use the quark-flavor basis mixing scheme \cite{quarkbasis},
\bea
\left(
	\begin{array}{c}
	\eta \\
	\eta^\prime
	\end{array}
\right)
=\left(
	\begin{array}{cc}
	{\rm cos} \, \phi & -{\rm sin} \, \phi \\
	{\rm sin} \, \phi & {\rm cos} \, \phi
	\end{array}
\right)
\left(
	\begin{array}{c}
	\eta_q \\
	\eta_s
	\end{array}
\right),
\label{eq:quarkbasis}
\eea
with $|\eta_q\rangle=\frac{1}{\sqrt{2}}(|u {\bar u} \rangle + |d {\bar d} \rangle$ and $|\eta_s \rangle = | s {\bar s} \rangle$.
The mixing angles in the two schemes are related via $\theta = \phi - {\rm arctan}\sqrt{2} \simeq \phi - 54.7^\circ$.
A two-mixing-angle scheme has also been suggested in the study of the mixing of decay constants \cite{twoangles}.

The mixing angle can neither be calculated from first principles in QCD nor measured directly -- it has to be determined phenomenologically.
There are a lot of studies on this subject  using different methods and a number of different processes, including various
decay processes involving the light pseudoscalar mesons \cite{decays,quarkbasis,twoangles,ChPT,smallratio,CEThomas07etal}.

One important source of information in determining the mixing angle is 
the transition processes, $\gamma \gamma^* \ra \eta, \ \etap$ for which the transition from factors (TFFs),
$F_{\eta \gamma}(Q^2)$ and $F_{\etap \gamma}(Q^2)$ with $Q^2$ being the virtuality of the off-mass-shell photon, are defined.
The usual procedure \cite{mixingfromTFFs} using the TFFs to evaluate the $\eta$-$\etap$ mixing angle is to calculate the $Q^2$ dependence
of these transition form factors and compare with the experimental data
which are given at a certain range of $Q^2$ \cite{CELLO,CLEO,BaBar_eta}.
However, theoretical calculations for the TFFs at finite $Q^2$ suffer sizable corrections and are sensitive to the non-perturbative model used for
the distribution amplitude of the mesons, which results in large uncertainties in determining  the mixing angle.

%A model-independent method of extracting the $\eta$-$\etap$ mixing parameters was proposed in \cite{CaoS_mixing99}.
Two analytical constraints on the $\eta$-$\etap$ mixing were obtained in \cite{CaoS_mixing99}
by considering the two-photon decays of $\eta$ and $\etap$ and
the asymptotic behavior of the $\eta$ and $\etap$ TFFs in the limit  $Q^2 \ra \infty$,
together with the fact that the asymptotic behavior of the meson TFFs is firmly predicted by QCD \cite{LepageB80}.
Newly available \babar\ data \cite{BaBar_eta} extend the measurements for the $\eta$ and $\etap$ TFFs
to  higher $Q^2$ and to a much larger range of $Q^2$,
and thus provide new information for the $\eta$ and $\etap$ TFFs at $Q^2 \ra \infty$.
At the same time experimental information on the masses and two-photon decay widths of mesons involved are improved considerably over the last decade.
These new experimental data shall have an impact on the determining of  the $\eta$-$\etap$ mixing parameters.

In this paper we extract the $Q^2 \ra \infty$ behavior of the $\eta$ and $\etap$ TFFs from the \babar\ data.
Using this new information and updated experimental data about the two-photon decays
$\eta \ra \gamma \gamma$ and $\eta^\prime \ra \gamma \gamma$ \cite{PDG2010},
we determine the $\eta$-$\etap$ mixing angle and the ratios of decay constants 
in the two mixing schemes [see Eqs.~(\ref{eq:eta81basis}) and (\ref{eq:quarkbasis})]
using the method of~\cite{CaoS_mixing99}.

The analytical expressions obtained in \cite{CaoS_mixing99} for the mixing angle $\theta$ and the ratio of the decay constants of
the $\eta_1$ and $\eta_8$ states $r=f_1/f_8$ are
\begin{widetext}
\bea
{\rm tan}\, \theta = \frac{ -(1+c^2)(\rho_1+\rho_2) 
	+\sqrt{(1+c^2)^2(\rho_1+\rho_2)^2
	+4 (c^2-\rho_1 \rho_2)(1-c^2\rho_1 \rho_2)} }
	{2(c^2 -\rho_1 \rho_2)},
\label{tantheta}
\eea
\bea
r = \frac{ (1+c^2)(\rho_1 -\rho_2) 
	+\sqrt{(1+c^2)^2(\rho_1 - \rho_2)^2+4 c^2(1+\rho_1 \rho_2)^2} }
	{2 c(1 +\rho_1 \rho_2)},
\label{rresult}
\eea
\end{widetext}
where $c=\sqrt{8}$ and
\bea
\rho_1&=&\left[
\frac{\Gamma_{\eta \ra \gamma \gamma}}
{\Gamma_{{\etap} \ra \gamma \gamma}}
\,\frac{m_{\etap}^3}{m_\eta^3}
\right]^{1/2},
\label{rho1} \\
\rho_2&=&\frac{F_{\eta \gamma}(Q^2 \ra \infty)}
{F_{\etap \gamma}(Q^2 \ra \infty)}.
\label{rho2}
\eea

One advantage of determining the mixing parameters from Eqs.~(\ref{tantheta})-(\ref{rho2}) is that both
the theoretical uncertainty incurred in calculating the TFFs at finite $Q^2$ and the experimental uncertainty
are minimized by considering the ratios of the decay widths for the two-photon decay processes
and the ratios of the transition form factors at large $Q^2$.

Furthermore, considering the ratio of the decay widths for the $\pi^0 \ra \gamma \gamma$ and 
$\eta \ra \gamma \gamma$ processes, we can also determine the ratios $f_8/f_\pi$ and $f_1/f_\pi$,
\bea
\frac{f_8}{f_\pi}&=&\rho_0
\left[
\frac{c_8}{c_\pi} \, {\rm cos} \,\theta
 - \frac{1}{r}\,\frac{c_1}{c_\pi}\, {\rm sin}\, \theta
\right],
\label{f8overpi} \\
\frac{f_1}{f_\pi}&=&\rho_0
\left[
\frac{c_8}{c_\pi} \,r \, {\rm cos} \, \theta 
- \frac{c_1}{c_\pi}\, {\rm sin}\, \theta
\right],
\label{f1overpi}
\eea
where $c_\pi=1$, $c_8=1/\sqrt{3}$, $c_1=2 \sqrt{2}/\sqrt{3}$, and
\bea
\rho_0=\left[
\frac{\Gamma_{\pi^0 \ra \gamma \gamma}}
{\Gamma_{{\eta} \ra \gamma \gamma}}
\,\frac{m_\eta^3}{m_{\pi^0}^3} \right]^{1/2}.
\label{rho0}
\eea

The above analysis can be easily applied to the quark-flavor basis mixing scheme [see Eq.~(\ref{eq:quarkbasis})] by replacing the parameters
$c=c_1/c_8$, $r=f_1/f_8$, $c_8$ and $c_1$ with $c^\prime=c_s/c_q=\sqrt{2}/5$, $r^\prime=f_s/f_q$, $c_q=5/3$ and $c_s=\sqrt{2}/3$,
respectively \cite{CaoS_mixing99}. 

The parameters $\rho_0$ and $\rho_1$ can be fixed
by using the masses and two-photon decay widths of $\pi^0$, $\eta$ and $\etap$.
We employ the data given by the 2010 Particle Data Group (PDG2010) \cite{PDG2010},
\bea
\Gamma_{\pi^0 \ra \gamma \gamma}&=& 7.74 \pm 0.46~{\rm eV}, \nonumber \\
\Gamma_{\eta \ra \gamma \gamma}&=& 0.510 \pm 0.026~{\rm keV}, \label{decaywidth}\\
\Gamma_{\etap \ra \gamma \gamma}&=& 4.28 \pm 0.19~{\rm keV}, \nonumber
\eea
\bea
m_{\pi^0} &=& 134.9766 \pm 0.0006~{\rm MeV}, \nonumber \\
m_\eta &=& 547.853 \pm 0.024~{\rm MeV}, \label{mass} \\
m_{\etap} &=& 957.78 \pm 0.06~{\rm MeV}. \nonumber
%PDG98
%m_{\pi^0} &=& 134.9764 \pm 0.0006 {\rm MeV}, \nonumber \\
%m_\eta &=& 547.30 \pm 0.12 {\rm MeV}, \label{mass} \\
%m_{\etap} &=& 957.78 \pm 0.14 {\rm MeV}. \nonumber
\eea

We will use the CLEO \cite{CLEO} and \babar\ \cite{BaBar_eta} data for the TFFs at large $Q^2$ to determine the parameter $\rho_2$.
The CLEO Collaboration \cite{CLEO} measured $F_{\eta \gamma}(Q^2)$ and $F_{\etap \gamma}(Q^2)$
in the $Q^2$ regions up to $20$ and $30$ GeV$^2$ respectively,
and presented the data in a mono-pole form proposed in \cite{sjbL81},
\bea
\left| F_{P \gamma}(Q^2) \right|^2= \frac{1}{(4 \pi \alpha)^2}
	\frac{ 64 \pi \Gamma_{P \ra \gamma \gamma}}{m_P^3}
	\frac{1}{\left(1+Q^2/\Lambda_P^2\right)^2}, \nonumber \\
\label{eq:FQexp}
\eea
where $\alpha\simeq 1/137$ is the QED fine coupling constant and $\Lambda_P$ is the pole mass parameter.
%\bea
%\Lambda^{\rm CLEO}_\eta &=& 774 \pm 11 \pm 16 \pm 22 \,{\rm MeV}, \nonumber \\
%\Lambda^{\rm CLEO}_{\etap} &=& 859 \pm 9 \pm 18 \pm 20 \,{\rm MeV}, \label{Lambda_CLEO}
%\eea
%In Eq. (\ref{Lambda_CLEO}) the first error represents statistical, the second error is systematic,
%and the third error comes from the uncertainty in the value of
%$\Gamma_{\eta \, (\etap) \ra \gamma \gamma}$.

The \babar\ Collaboration \cite{BaBar_eta} recently measured the $\eta$-photon and $\etap$-photon transition form factors 
in the $Q^2$ range from $4$ to $40$ GeV$^2$.
The results were not presented in the mono-pole form [Eq. (\ref{eq:FQexp})], partially 
because their results for the pion-photon transition from factor exhibit a very quick growth for $Q^2>15$ GeV$^2$ \cite{BaBarpi0},
which is very hard to explain in QCD \cite{BaBarExplanations}.
However, this trend of fast growth is noticeably missing from the \babar\ data for the $\eta$-photon and $\eta^\prime$-photon transition form factors,
and thus the \babar\ data for the $\eta$-photon and $\eta^\prime$-photon transition form factors are consistent with perturbative QCD calculations for the form factors
and shall be described with the mono-pole form as given by Eq.~(\ref{eq:FQexp}).

We use QCD-motivated mono-pole form Eq.~(\ref{eq:FQexp}) to fit experimental data.
The values of $\Lambda_\eta$ and $\Lambda_\etap$ 
in Eq.~(\ref{eq:FQexp}) determined using the CLEO data, \babar\ data, and the combined data are presented in Table~{\ref{Table:Lambda}.
We have combined the statistical and systematic errors for the CLEO data in quadrature since the \babar\ data are presented with only combined errors.
The values of $\chi^2/$d.o.f given in the table provide further justification for the use of Eq.~(\ref{eq:FQexp}) in describing these data.
The values of $\Lambda_\eta$ and $\Lambda_\etap$ determined with the CLEO and \babar\ data  agree within their uncertainties,
but the \babar\ data greatly improve the accuracy in determining the values of $\Lambda_\eta$ and $\Lambda_\etap$.
Using the combined data in the fitting changes the results slightly.

The parameter $\Lambda_P$ has a natural explanation as the pole mass of vector meson in the vector meson dominated model for the TFFs.
The values we obtained, $\Lambda_\eta \sim 780$ MeV and $\Lambda_\etaÕ \sim 860$ MeV, are very close to the masses of $\rho$ (770 MeV) and $K^*$ (890 MeV).

\begin{table*}%[H] add [H] placement to break table across pages
\caption{$\Lambda_P$ and $\chi^2/$d.o.f in fitting the data for the TFFs with Eq.~(\ref{eq:FQexp}).
%\caption{The values of $\Lambda_P$ in Eq.~(\ref{eq:FQexp}) determined using the CLEO data, \babar\ data, and the combined data.
\label{Table:Lambda}}
%{\vskip 0.3cm}
\begin{tabular}{|c|c|c|c|c|}
\hline
\multirow{2}{*}{} & \multicolumn{2}{c|}{$\eta$} & \multicolumn{2}{c|}{$\eta^\prime$}  \\ \cline{2-5}
 & $\Lambda_\eta$ (MeV) & $\chi^2/$d.o.f  & $\Lambda_{\eta^\prime}$ (MeV) & $\chi^2/$d.o.f  \\ \hline
CLEO   & 775 $\pm$ 12  & 0.95 &   $856 \pm 10$ & 0.88 \\ \hline
\babar\  & 787 $\pm$ 7   & 0.99 &  $861 \pm 4$ & 1.04  \\ \hline
CLEO+\babar\ & 784 $\pm$ 6 & 0.96 & $849 \pm 6$ & 0.88 \\ \hline
 \end{tabular}
 %\end{ruledtabular}
 \end{table*}
%\begin{table*}%[H] add [H] placement to break table across pages
%\caption{Fitting the data with dipole form factor).
%\caption{The values of $\Lambda_P$ in Eq.~(\ref{eq:FQexp}) determined using the CLEO data, \babar\ data, and the combined data.
%\label{Table:Lambda}}
%{\vskip 0.3cm}
%\begin{tabular}{|c|c|c|c|c|}
%\hline
%\multirow{2}{*}{} & \multicolumn{2}{c|}{$\eta$} & \multicolumn{2}{c|}{$\eta^\prime$}  \\ \cline{2-5}
% & $\Lambda_\eta$ (MeV) & $\chi^2/$d.o.f  & $\Lambda_{\eta^\prime}$ (MeV) & $\chi^2/$d.o.f  \\ \hline
%CLEO   & 971 $\pm$ 8  & 0.77 &   $998 \pm 6$ & 2.47 \\ \hline
%\babar\  & 920 $\pm$ 4   & 1.35 &  $961 \pm 2$ & 2.59  \\ \hline
%CLEO+\babar\ & 928 $\pm$ 4 & 1.92 & $977 \pm 3$ & 3.21 \\ \hline
 %\end{tabular}
 %\end{ruledtabular}
 %\end{table*}
\begin{table*}%[H] add [H] placement to break table across pages
\caption{The mixing parameters determined for the $\eta_8$-$\eta_1$ mixing scheme.
% for $\rho_2$ given in Eq.~(\ref{rho2}).
\label{Table_mixing81}}
\begin{tabular}{|c|c|c|c|c|}
\hline
 & $\theta$ &  $f_1/f_8$ & $f_8/f_\pi$ &$f_1/f_\pi$ \\ \hline
CLEO   & $-16.26 \pm 0.86 $   & 1.162 $\pm$ 0.053   & 0.955 $\pm$ 0.042   & 1.109 $\pm$ 0.053   \\ \hline
\babar\  & $-16.54 \pm 0.71 $& 1.146 $\pm$ 0.045 & 0.966 $\pm$ 0.041& 1.107 $\pm$ 0.050  \\ \hline
CLEO+\babar\ & $-16.84 \pm 0.72 $& 1.128 $\pm$ 0.044 & 0.979 $\pm$ 0.042 & 1.105 $\pm$ 0.050  \\ \hline
 \end{tabular}
 %\end{ruledtabular}
 \end{table*}
\begin{table*}%[H] add [H] placement to break table across pages
\caption{The mixing parameters determined for the quark-flavor basis mixing scheme.
% for $\rho_2$ given in Eq.~(\ref{rho2}).
\label{Table_mixingquark}}
\begin{tabular}{|c|c|c|c|c|}
\hline
 & $\phi$ &  $f_s/f_q$ & $f_q/f_\pi$ &$f_s/f_\pi$ \\ \hline
CLEO   & $38.11 \pm 0.79 $   & 1.197 $\pm$ 0.065   & 1.076 $\pm$ 0.044   & 1.29 $\pm$ 0.10   \\ \hline
\babar\  & $37.90 \pm 0.70 $& 1.177 $\pm$ 0.054     & 1.077 $\pm$ 0.044   & 1.268 $\pm$ 0.088  \\ \hline
CLEO+\babar\ & $37.66 \pm 0.70 $& 1.156 $\pm$ 0.054 & 1.078 $\pm$ 0.044 & 1.246 $\pm$ 0.087  \\ \hline
 \end{tabular}
 %\end{ruledtabular}
 \end{table*}

%\bea
%\Lambda^{\rm BaBar}_\eta &=& 774 \pm 11 \pm 16 \pm 22 \,{\rm MeV}, \nonumber \\
%\Lambda^{\rm BaBar}_{\etap} &=& 859 \pm 9 \pm 18 \pm 20 \,{\rm MeV}. \label{Lambda_BaBar}
%\eea

The results for the mixing angle and decay constants determined using the CLEO data, \babar\  data, and the combined data,
together with the two-photon decay widths, are presented in Tables \ref{Table_mixing81} and \ref{Table_mixingquark}
for the $\eta_8$-$\eta_1$ mixing scheme and quark-flavor basis mixing scheme, respectively.
The mixing angle obtained in this work, $\phi \simeq 37^\circ \sim 38^\circ$ is slightly smaller than the central value of $39.8^\circ$ 
obtained in \cite{CaoS_mixing99}. This is mainly due to an increase in the estimation for the $\Gamma_{\eta \ra \gamma\gamma}$
by the 2010 Particle Data Group. This increase also affects the results for the ratios of decay constants slightly.
The uncertainties for the mixing angles and the ratios $f_1/f_8$ and $f_s/f_q$ obtained in this work are considerably smaller than 
that given in \cite{CaoS_mixing99} due to the new more accurate experimental data for the meson masses, the two-photon decay widths,
and the meson-photon transition form factors.
The uncertainties for the other ratios of decay constants, $f_8/f_\pi$ and $f_1/f_\pi$ in the $\eta_8$-$\eta_1$ mixing scheme
and $f_q/f_\pi$ and $f_s/f_\pi$ in the quark-flavor basis mixing scheme,
are also generally smaller than that estimated in \cite{CaoS_mixing99}.

Our results for the mixing angle are in agreement with recent results of $\phi \simeq 37^\circ \sim 42^\circ$ 
obtained with other methods \cite{CEThomas07etal}.
The value of $f_8/f_\pi$ obtained in this work is smaller than that obtained with Chiral Perturbation Theory (ChPT) at next-to-leading order \cite{ChPT}
and some phenomenological analyses \cite{CEThomas07etal},
but is larger than the result reported in \cite{smallratio}.
We note that the ChPT result may be alerted by higher order corrections.
As it has been pointed out in \cite{CaoS_mixing99}, in the previous studies either the questionable assumption that the decay constants and the particle states share 
the same mixing scheme or two mixing-angle scheme was adopted. The relations between the mixing parameters involved
in the two-mixing-angle scheme and that appear in our model remain to be further studied.

In summary, understanding the composition of the light pseudoscalar mesons $\eta$ and $\etap$ is of great importance in the
study of many hadron processes involved these mesons.
Employing the two analytical constrains on the $\eta$-$\etap$ mixing proposed by us in a previous work, 
we extracted the $\eta$-$\eta^\prime$ mixing angle and the ratios of decay constants in two widely-used mixing schemes
using recently available \babar\ measurements on
the $\eta$-photon and $\eta^\prime$-photon transition form factors and more accurate experimental data for the masses and two-photon decay widths of
the light pseduoscalar mesons. 

%\section*{Acknowledgements}
%\begin{acknowledgments}
F.G. thanks X.-H. Guo for his hospitality at Beijing Normal University where part of this work was done.
%\end{acknowledgments}

%\bibliography{PionTFF7_PRD.bib}

\end{document}